# ANGULAR DISTRIBUTIONS OF THE PARTICLES EMITTED IN Kr (AT 0.95 A GeV) AND Au (AT 10.7 A GeV) EMULSION REACTIONS.

M.K. Suleymanov[1,2*], O.B. Abdinov[3], R.M. Aliyev[3], F.M. Aliyev[3], Ya.H. Huseynaliyev[2,5],
E.U. Khan[2], A. Kravčáková[4], E.I. Shahaliev[1], S. Vokál[1,4], A.S. Vodopianov[1].

[1)] JINR, Dubna; [2)] CIIT, Islamabad; [3)] Physics Institute of AS, Baku;
[4)] University of P.J. Šafárik, Košice;
[5)] Sumgayit State University, Azerbaijan.
\* E-mail: mais@jinr.ru

**Abstract**

We will discuss the experimental results of the behavior of the angular distributions of slow particles emitted in hadron-nuclear and nuclear-nuclear interactions at relativistic energies.

## 1. Introduction

Searching for the signals on the formation and decay of cluster formation [1, 2], formation of the fireballs [3] and other intermediate baryon objects could be considered as a very important tool to reach the new states and phases of nuclear matter.

Formations of these objects in the interactions of the hadrons and nuclei with nuclei have to influence on the characteristics of secondary particles and this lead to changing the behavior of the ones. By studying these changing, it is possible to get some important information about the dynamics of the interactions. There are many papers that predict that the angular distributions of fragments could have some special structure as a result of the formation and decay of some intermediate formations (for example see [4]). There are only a few experimental results [5-8] that could be considered as some confirmation of it. These results mainly belong to the interactions of hadrons and light nuclei (we are going to discuss its below). But there is not any information about the same structure for the heavy nuclei interactions. So the main goal of our work is to analyze the angular distributions of secondary slow particles emitted in such interactions. We have used the experimental data on Kr+Em - reaction at 0.95 A GeV [9] and Au+Em - reaction at 10.6 A GeV [10].

This information is necessary to understand the dynamic of regime change on the behavior of characteristics of the particles as a function of centrality [11]. It is possible that the reason of the structure in angular distributions of slow particles in central collisions is the formation and decay of the nuclear cluster through the percolation mechanisms (for example see [12]). We, therefore, think about percolation mechanism because the structure was observed mainly in hadron-nucleus and light nucleus-nucleus central collisions. It will be very perspective to continue these investigations because now there are some very interesting ideas connected with the formation and decay intermediate baryon objects through the percolation mechanism [13]. Let us consider some of them.

The paper [14] discusses that deconfinement is expected when the density of quarks and gluons becomes so high that it no longer makes sense to partition them into color-neutral hadrons, since these would strongly overlap. Instead we have clusters much larger than hadrons, within which color is not confined; deconfinement is thus related to cluster formation. This is the central topic of percolation theory, and hence a connection between percolation and deconfinement seems very likely [15]. Observing of the effects connected with formation and decay of the percolation clusters in heavy ion collisions at ultrarelativistic energies could be the first step for getting the information of the onset stage of deconfinement.

Another idea is connected with nucleon coalescence effect in heavy ion interactions [16]. There is a very great chance that the effect of the light nuclei emission in heavy ion collisions are to be one of the accompany effects of percolation cluster formation and decay. Since the probability of coalescence of a particular nuclear system depends on the properties of the hadronic system that formed as a result of the collision, so it could be expected that probability of nucleon coalescence process could increase with growing nuclear matter density, with the formation of percolation cluster.

Light nuclei are fairly large objects compared to simple hadrons and their binding energies are small compared to freeze out temperatures, which are on the order of 100 MeV. These light clusters are, therefore, not expected to survive through the high-density stages of the collision. The light nuclei observed in the experiment are formed and emitted near freeze-out, and they mainly carry information about this late stage of the collision. This is evident from the simple nucleon coalescence model [17]. Existing of the light nuclei produced as a result of coalescence has to change the behavior of the centrality dependences of light nuclei yields. It is expected the regime change on the behavior of light nuclei yields as a centrality of collisions.

To fix the baryon density of nuclear matter, the centrality experiments are usually used.

## 2. Angular distribution of the fragments

As we mentioned above, there are only a few experimental results [5-8], which could be considered as possible confirmation of the existing some structure on the angular distribution of secondary particles mainly for the emitted protons and for the slowest fragments. Let us consider them.

### 2.1. Hadron-nuclear interactions

In paper [5] it was obtained that the angular distributions of protons emitted in $\pi^-$ $^{12}$C-interaction (at 40 GeV/c) with total disintegration of nuclei (or central collisions) have some structure, pick at angles close to $60^0$ (see Fig.1). This result was confirmed by the data, which were obtained at the investigation of the angular distributions of protons emitted in $\pi^{-12}$C-interaction with total disintegration of nuclei at 5 GeV/c [6] (see Fig.2). In both works were considered protons with momentum less than 1 GeV/c.

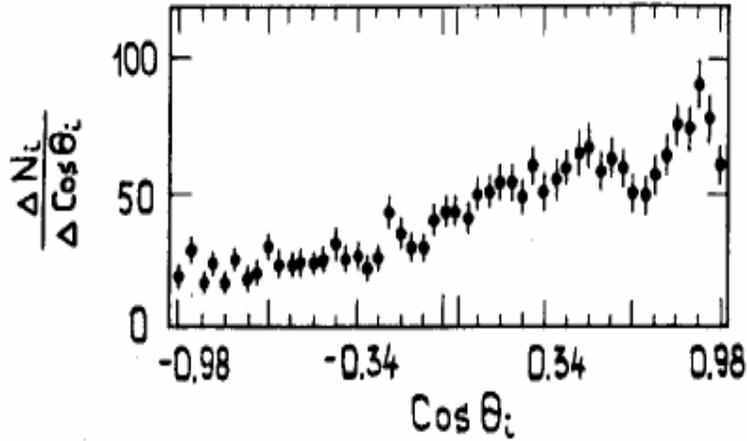

Fig.1. The angular distributions of protons emitted in $\pi^{-12}$C-interaction (at 40 GeV/c) with total disintegration of nuclei.

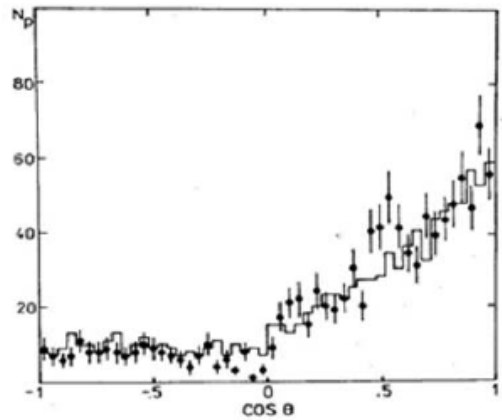

Fig.2. The angular distributions of protons emitted in $\pi^{-12}$C-interaction (at 5 GeV/c) with total disintegration of nuclei.

### 2.2. Interactions of light and middle mass nuclei with emulsion ones

In Fig. 3 the angular distributions for the slow protons emitted in the central He+Em- (at 2.1 A GeV), O+Em - (2.1 A GeV) and Ar+Em - (1.8 A GeV) collisions are shown. The data were taken from paper [7]. The some wide structure was observed in these distributions.

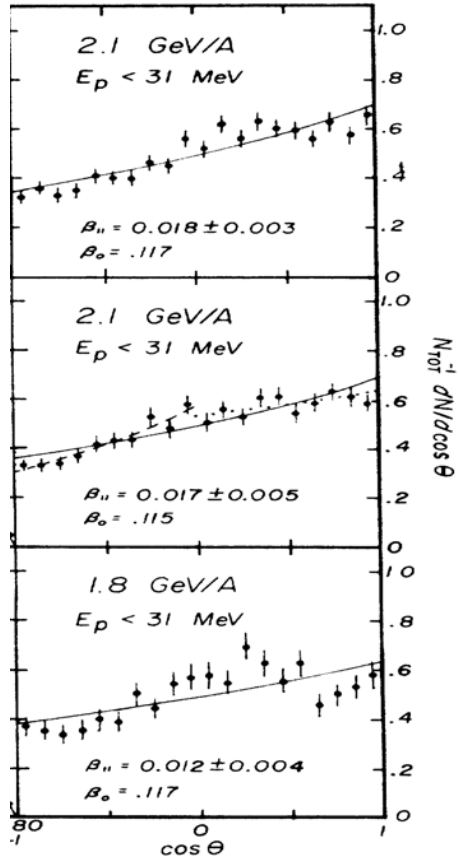

Fig.3. The angular distribution for slow protons emitted in the central He+Em-( at 2.1 A GeV), O+Em - (2.1 A GeV) and Ar+Em - (1.8 A GeV) collisions.

Almost similar but wider structures have been observed on the angular distributions of the b-particles emitted in the Ne+Em reactions at 4.1 A GeV (see Fig.4-5) [8]. They commented the structure become cleaner in central collisions.

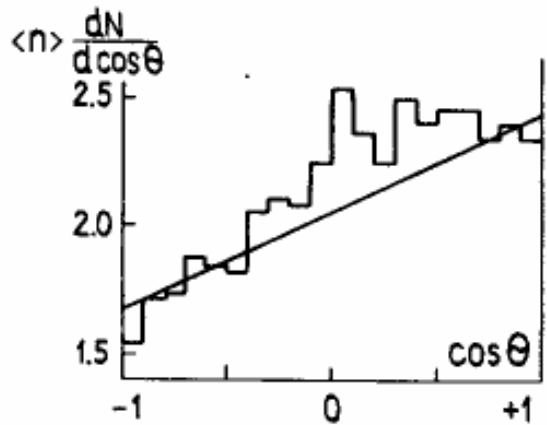

Fig.4. The angular distributions of the b-particles emitted in the Ne+Em reactions at 4.1 A GeV.

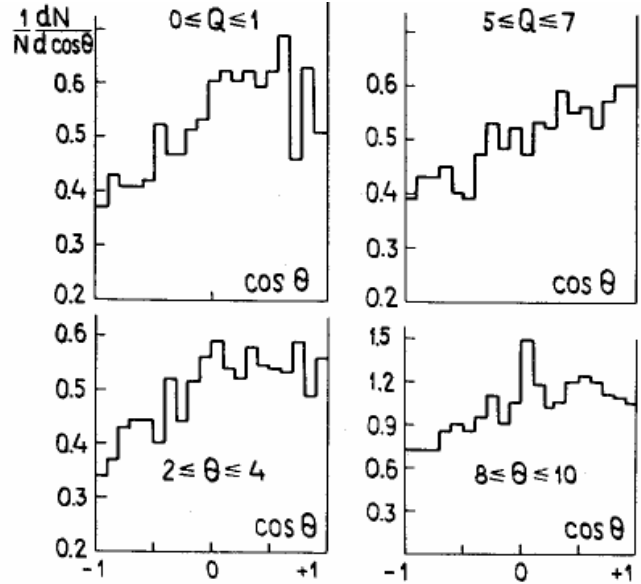

Fig.5. The angular distributions of the b-particles emitted in the Ne+Em reactions (with different centralities Q, the condition $0 \leq Q \leq 1$ corresponds to the most central event and $8 \leq Q \leq 10$ to peripheral ones) at 4.1 A GeV.

The scattering and rescattering effects could explain the peak that is in angular distribution of b-particles emitted in peripheral Ne+Em collisions.

### 3. The results of the heavy nuclei interaction

So we could see above that there are experimental results, which demonstrated the existing of some structures on the angular distributions of protons and b-particles emitted in the interactions of hadrons, light and middle mass nuclei with nuclear targets. But there is not any information about same structure for the interaction of heavy nuclei with nuclear targets. Getting this information was the main goal of our investigation and to reach it the experimental data on Kr+Em - reaction at 0.95 A GeV [9] and Au+Em - reaction at 10.6 A GeV [10] are processed by us. These data were obtained by the EMU01 Collaborations [9,10] using nuclear beams of the AGS BNL.

The angular distribution of slow particles was presented by the EMU01 Collaboration (for example see [18]). But in our investigation the angular distribution of the slow fragments were considered separately for:

1) all events and the events with a number of $N_h \geq 8$ ($N_h$ is a number of h-particles in emulsion experiments, this criterion have been used to separate the heavy nuclei interaction by the Collaboration);
2) the peripheral and central collisions.

The experimental data have been compared with data coming from the cascade evaporation model (CEM, see [19]).

Fig 6 a-d show the angular distributions of the b- and g – particles emitted in the Kr + Em - reactions at 0.95 A GeV. The results are shown separately for all events (Fig.6a, b) and the events with $N_h \geq 8$ (Fig.6c, d). The results coming from the CEM have also been shown in these figures. We cannot see any structure on these distributions that observed in Figs.1-5.

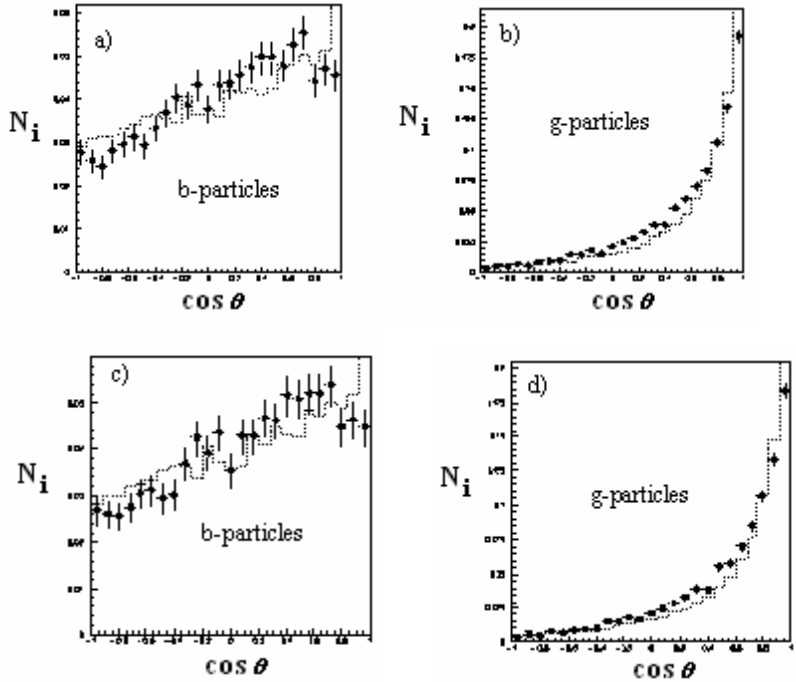

Fig.6 a-d. The angular distributions of the particles emitted in Kr+Em-reactions at the energy 0.95 A GeV (black circles are experimental data): a) b-particles and b) g-particles emitted in the all events; c) b-particles and d) g-particles emitted in the events with $N_h \geq 8$. The histograms are the results coming from the CEM.

In Fig 7a-d are shown the angular distributions of the b- and g – particles emitted in the Au + Em - reactions at 10.7 A GeV. As in Fig. 6 a-d, the results are demonstrated separately for all events (Fig.7a-b) and the events with $N_h \geq 8$ (Fig.7c-d).

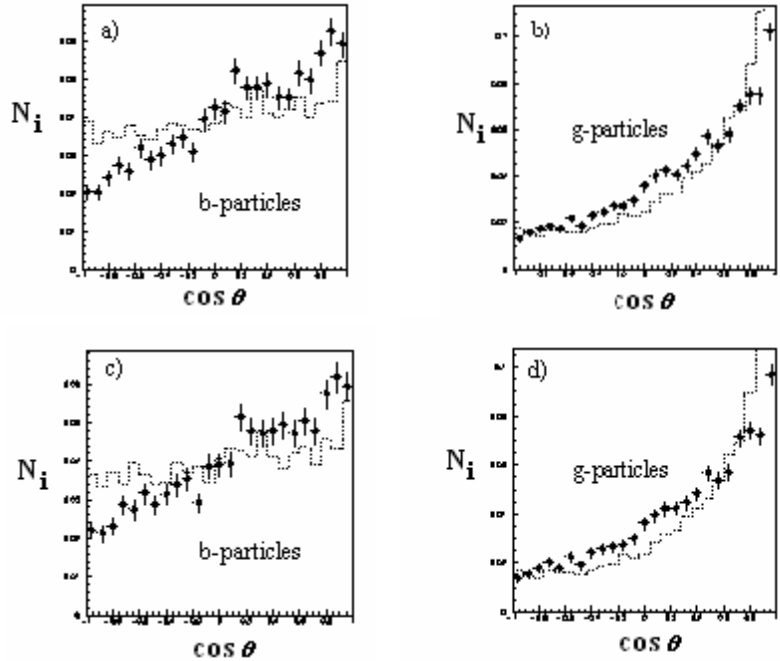

Fig.7 a-d. The angular distributions of the particles emitted in Au+Em-reactions at the energy 10.7 A GeV (black circles are experimental data): a) b-particles and b) g-particles emitted in the all events; c) b-particles and d) g-particles emitted in the events with $N_h \geq 8$. The histograms are the results coming from the CEM.

These figures show some structure in the angular distributions for b- and g-particles. CEM could not describe completely the angular distributions of b-particles. We could also say that there is some wide structure in the angular distributions of these particles in the events with $N_h \geq 8$. We believe that this structure could be connected with elastic scattering of the internuclei nucleons. That is why this structure is seen more cleanly in all the events.

There is some systematic deviation for the experimental data obtained for g-particles from the ones coming from the CEM.

In the Figs.8 a-d are demonstrated the angular distributions of the b-, and g – particles emitted in the Kr + Em - reactions at 0.95 A GeV. The results are shown separately for the central collisions (Figs.8a-b) and for the peripheral ones (Figs.8c-d). To select the central collisions, we used the criteria $Ng \geq 20$ which was obtained in paper [20].

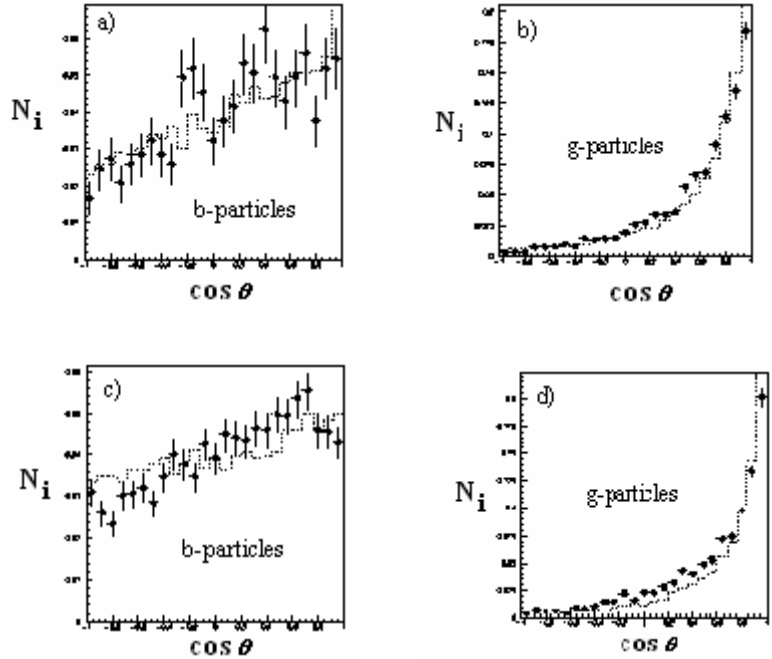

Fig.8 a-d. The angular distributions of the particles emitted in Kr+Em-reactions at the energy 0.95 AGeV (black circles are experimental data): a) b-particles and b) g- particles emitted in the central collisions; c) b-particles and d) g-particles emitted in the peripheral collisions. The histograms are the results coming from the CEM.

We can see that there is not any special behavior for these distributions, which could not be described by CEM as well as in Fig.6 a-d.

In the Fig.9a-d are demonstrated the angular distributions of the b- and g – particles emitted in the Au + Em - reactions at 10.7 A GeV. The results are shown separately for the central collisions (Fig.9a-b) and for the peripheral ones (Fig. 9c-d). To select the central collisions, we used the criteria $N_F \geq 20$ though in paper [20]. It was shown that the central Au+Em-events (at 10.7 A GeV) should be selected using the criteria $N_F \geq 40$. We, therefore, did not use last criteria because a number of events with $N_F \geq 40$ are very small.

We can see that in peripheral collisions for b-particles, the structure becomes cleaner. It could mean that these structures are as a result of nucleon elastic scattering. It is seen that the structure gets weak and almost disappear in central collisions.

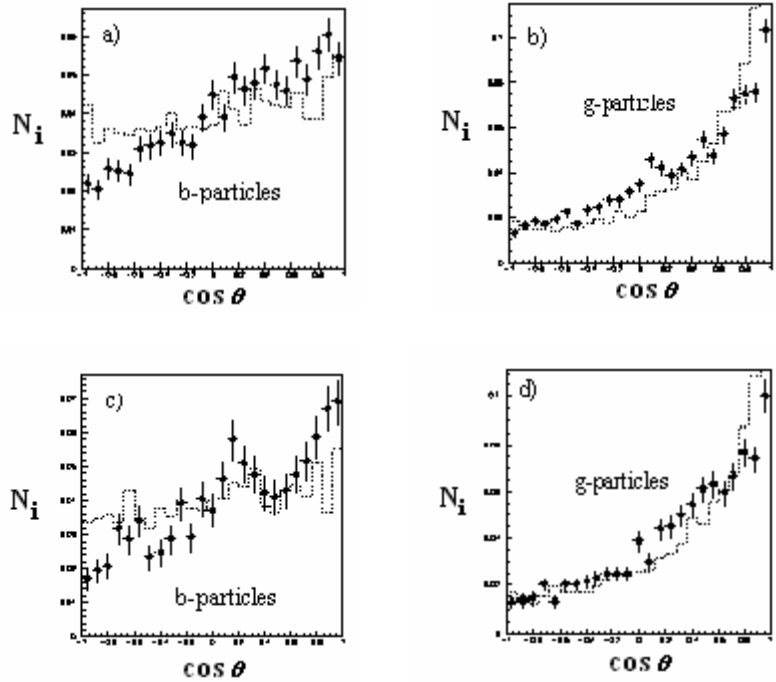

Fig.9 a-d. The angular distributions of the particles emitted in
Au+Em-reactions at the energy 10.7 A GeV (black circles are experimental
data): a) b-particles and b) g-particles emitted in the central collisions; c) b-
particles and d) g-particles emitted in the peripheral collisions. The histograms
are the results coming from the CEM.

## 4. Discussion the results and summary

So we could see that the angular distributions of b-and g-particles emitted in Au+Em-reactions as well as in the Kr+Em-reactions don't contain any special structure which could not be described using the usual mechanisms of the interaction such as cascade evaporation, elastic scattering.

If we will turn to the results which are represented in paragraphs 2.1, 2.2 and will compare them with the result obtained for Kr+Em and Au+Em, so we could see that with increasing the mass of the projectile particle the angular distribution of slow particles change and the structure which was mentioned above almost disappeared.

In paper [20] the behavior of the distributions of the target (b -, g - and h - particles) and the projectile fragments (F -particles) emitted in Kr + Em - (at energies 0.95 A GeV) and in Au + Em- reactions (at energies 0.95 A GeV) have been studied. Comparing the experimental data with ones coming from the cascade-evaporation model [19], it has been concluded that the formation of cluster could sufficiently influence the characteristics of nuclear fragments.

In paper [2] the results of intranuclear cascade calculations (ideal gas with two-body collisions and no mean field), complemented by a simple percolation procedure, are compared with experimental data on protons and light nuclear fragments ($d$, $t$, $^3He$, and $^4He$) measured in 400 and 800 A MeV Ne+Nb collisions using a large solid angle detector. The model reproduces quite well global experimental observables like nuclear fragment multiplicity distributions or production cross-sections, and nuclear fragment to proton ratios.

The results that were presented in papers [2,20] demonstrated that to describe the multiplicity distribution of slow particles, it is necessary to use special mechanisms as well as percolation one. But for angular distributions, the influence of any other mechanisms than cascade evaporation ones and elastic scattering mechanisms appears to be negligible small.

This result could be explained by the following way: during the interaction of the projectile with nuclear target with increasing the mass of the projectile; a number of secondary interactions are growths as well as a number of nucleon-nucleon elastic scatterings and rescattering events. These effects could lead to the disintegration of any intermediate formations as well as clusters, decreasing their influence (expect to multiplicity distribution) on the characteristics of the emitted particles.

In ultrarelativistic ion interactions, it was also observed the decreasing (or suppression) some effects, which were observed for light nuclear collisions. So at CERN SPS energy the NA50 [21] experiments observed the $J/\psi$ suppression in Pb+Pb interaction at 158 GeV in comparison with the light nuclear collisions. At BNL RHIC energy the Star [22] and Phenix [23] experiments observe the suppression of high $p_t$ particles. It is supposed that the reason of these effects could be the formation the new phase of strongly interacting matter (for example Quark Gluon Plasma [24]). May be these results point up to that to get more true information about the intermediate nuclear systems more easy way could be to use the data of hadron-nuclear and light nuclear-nuclear interactions.

## Acknowledgements

We are indeed indebted for the support extended to us by the Agency for Science of the Ministry for Education of the Slovak Republic (Grant VEGA 1/2007/05), the Higher Education Commission of the Islamic Republic of Pakistan (Grant N1-28/HEC/HDR/2006) , CIIT (Islamabad) and JINR (Dubna).

## References

1. M. Petrovici et al. Phys. Rev. Lett. 74, N25 (1995); W.A. Friedman. Phys.Rev.C42, N2 (1990) pp.667-673; W. Reisdorf et al. Phys. Lett. B 595 (2004) pp.118–126; J. B. Elliott et al. E-print:nucl-ex/0002006, 2000; nucl-ex/0002004, 2000
2. G. Montarou et al. Phys. Rev. C 47:2764-2781,1993
3. D.Pelte et al.E-print:nucl-ex/9704009,1997
4. A. Agnese, M. La Camera, A. Wataghin (Genoa U.). Nuovo Cim.A59: 71-80,1969.
5. A.I.Anoshin et al.Yad.Fiz.33: 164(1981)
6. O.B.Abdinov et al. Preprint JINR, 1-80-859, Dubna (1980).
7. H.H. Hecman et al. Phys.Rev. C17, N5(1978) .
8. N.P. Andreeva et al. Yad.Fiz.45:123-131,1987
9. S.A.Krasnov et al., Czech. J. Phys. **46**, No.6, 531 (1996)

10. M.I. Adamovich et al., Phys. Lett. **B 352** (1995).
11. M.K. Suleymanov et al. Nucl.Phys.A734S: pp. E104-E107, 2004.
12. C. Volant et al. Nuclear Physics A734 (2004) 545-548
13. H. Satz. E-print: hep-ph/0212046 ,2002; C. Pajares.E-print: hep-ph/0501125, 2005; J. Brzychczyk. E-print: nucl-th/0407008.
14. H. Satz. ArXiv: hep-ph/0007069 v1 7 Jul 2000
15. H.Satz, Nucl.Phys.A642 (1998) 130c.;G. Baym, Physica 96 A (1979) 131.; T. Celik, F. Karsch and H. Satz, Phys. Lett. 97 B (1980) 128.
16. S.S. Adler et al.E-prin: nucl-ex/0406004 , 2004; Zhangbu Xu for the E864 Collaboration.E-prin: nucl-ex/9909012; S. Albergo, Phys. Rev. C, v. 65, 034907; I.G. Bearden et al. European Physical Journal C. The paper number is EPJC011124.
17. L.P. Csernai and J.I. Kapusta, Phys. Rep. **131**, 223 (1986); A.Z. Mekjian, Phys. Rev. **C17**, 1051 (1978).
18. M.I. Adamovich et al. Nucl.Phys.A 590 (1995), pp. 597-600; M.I. Adamovich et al. Z.Phys.A358 (1997), pp.337-351.
19. G.J. Musulmanbekov, Proc. of the 11th EMU01 Collaboration Meeting, Dubna, Russia, (1992); Proc. 11th Int. Symp. on High Energy Spin Phys., Bloomington, (1994); AIP Conf. Proc. **343**, 428 (1995).
20. O.B. Abdinov et al. Journal of Bulletin of the Russian Academy of Sciences. Physics, vol.70, N 5 (2006): e-Print Archive: hep-ex/0503032]
21. M.C Abreu et al. Phys.Let. B 450, p.456(1999); M.C Abreu et al. Phys.Let. B 410, p.337(1997); M.C Abreu et al. Phys.Let. B 410, p.327(1997);
22. J. Adams et al. Phys. Rev. Lett. **91** (2003) 072304
23. S.S. Adler et al. E-print: nucl-ex/0611007